\documentclass[nojss
]{jss}

\usepackage{orcidlink,thumbpdf,lmodern}

\usepackage[utf8]{inputenc}

\author{
Benjamin D. Youngman~\orcidlink{0000-0003-0215-8189}\\Department of
Mathematics and Statistics\\
University of Exeter
}
\title{\pkg{deform}: An R Package for Nonstationary Spatial Gaussian
Process Models by Deformations and Dimension Expansion}

\Plainauthor{Benjamin D. Youngman}
\Plaintitle{deform: An R Package for Nonstationary Spatial Gaussian
Process Models by Deformations and Dimension Expansion}
\Shorttitle{\pkg{deform}: Nonstationary Spatial Gaussian Process Models
in R}

\Abstract{
Gaussian processes (GP) are a popular and powerful tool for spatial
modeling of data, especially data that quantify environmental processes.
However, in stationary form, whether covariance is isotropic or
anisotropic, GPs may lack the flexibility to capture dependence across a
continuous spatial process, especially across a large domain. The
\pkg{deform} package aims to provide users with user-friendly
\proglang{R} functions for the fitting and visualization of
nonstationary spatial GPs. Users can choose to capture nonstationarity
with either the spatial deformation approach of \cite{sampson1992} or
the dimension expansion approach of \cite{bornn2012}. Thin plate
regression splines are used for both approaches to bring transformations
of locations to give a new set of locations that bring isotropic
covariance. Fitted models in \pkg{deform} can be used to predict these
new locations and to simulate nonstationary GPs for an arbitrary set of
locations.
}

\Plainkeywords{Gaussian process, geostatistics, spatial
deformation, dimension expansion, thin plate regression spline, R}

\Address{
    Benjamin D. Youngman\\
    Department of Mathematics and Statistics\\
University of Exeter\\
    Laver Building, North Park Road\\
Exeter, EX4 4QE, UK\\
  E-mail: \email{b.youngman@exeter.ac.uk}\\
  URL: \url{https://byoungman.github.io/}\\~\\
  }

\usepackage{amsmath, mathtools, amssymb}   \DeclarePairedDelimiter{\norm}{\lVert}{\rVert}

\begin{document}

\section{Introduction} \label{S:intro}

When modeling spatial processes, it may be inappropriate to assume that
dependence is both stationary and isotropic, especially when dealing
with large domains. For example, when modeling US rainfall, we might
expect a different dependence structure over mountainous regions from
over plains. This work allows nonstationarity in dependence by deforming
the space on which a process is typically defined to one in which
stationarity and isotropy are more appropriate. Formally consider
\(\boldsymbol{x} \in \mathbb{R}^p\) and a mapping
\(\boldsymbol{g} : \mathbb{R}^p \mapsto \mathbb{R}^q\) such that
\(\boldsymbol{x}^* = \boldsymbol{g}(\boldsymbol{x})\) for
\(\boldsymbol{x}^* \in \mathbb{R}^q\). \cite{sampson1992} introduced
this approach in the context of spatial modeling, so that \(p=q=2\):
then \(\boldsymbol{x} \in \mathbb{R}^2\), a coordinate in geographic
space (henceforth \(G\)-space), is mapped to
\(\boldsymbol{x}^* \in \mathbb{R}^2\), a coordinate in dispersion space
(henceforth \(D\)-space), where
\(\boldsymbol{x}^* = \boldsymbol{g}(\boldsymbol{x})\) for
\(\boldsymbol{g} : \mathbb{R}^2 \mapsto \mathbb{R}^2\). Stationarity and
isotropy are assumed for \(D\)-space. The remainder of this work focuses
on \(\boldsymbol{x} \in \mathbb{R}^2\).

\cite{sampson1992} represent \(\boldsymbol{g}\) as a pair of thin plate
splines, which are fitted to coordinates obtained through non-metric
multidimensional scaling (NMDS). Subsequent works by \cite{damian2001}
and \cite{schmidt2003} adopt a Bayesian approach to inference and assume
that \(\boldsymbol{g}\) is a random function and data are realizations
of a Gaussian process (GP); i.e.,
\begin{equation} \label{gp} Y_t(\boldsymbol{x}) \mid \boldsymbol{g} \sim GP\big(\mu(\boldsymbol{x}), v\big(\boldsymbol{g} (\boldsymbol{x}), .\big)\big)\end{equation}
for time \(t=1, \ldots, n\), location
\(\boldsymbol{x} \in \mathbb{R}^2\), and mean and covariance functions
\(\mu\) and \(v\), respectively. For a fixed set of locations,
\(\boldsymbol{x}_1, \ldots, \boldsymbol{x}_m\), this allows inference to
be based on the likelihood
\begin{equation}L(\boldsymbol{\Sigma}) = |2 \pi \boldsymbol{\Sigma}|^{-(n - 1) / 2} \exp\Big(-\dfrac{n}{2}\text{tr}(\boldsymbol{\Sigma}^{-1} \boldsymbol{V})\Big),\label{gplik}\end{equation}
where
\(\boldsymbol{V} = (n - 1)^{-1}\sum_{t=1}^n(\boldsymbol{y}_t - \hat{\boldsymbol{\mu}})(\boldsymbol{y}_t - \hat{\boldsymbol{\mu}})^\text{T}\)
with
\(\boldsymbol{y}_t = (y_t(\boldsymbol{x}_1), \ldots, y_t(\boldsymbol{x}_m))\),
\(m \times m\) matrix \(\boldsymbol{\Sigma}\) has \((i, j)\)th element
\(\Sigma_{i, j} = v(\boldsymbol{g} (\boldsymbol{x}_i), \boldsymbol{g} (\boldsymbol{x}_j))\)
and \(\hat{\boldsymbol{\mu}} = (\hat \mu_1, \ldots, \hat \mu_m)\) with
\(\hat \mu_i = n^{-1}\sum_{t=1}^n y_t(\boldsymbol{x}_i)\).
\cite{damian2001} and \cite{schmidt2003} place thin plate spline and
Gaussian process priors on \(\boldsymbol{g}\), respectively, and use
Markov chain Monte Carlo to sample from the posterior distribution of
\(\boldsymbol{g}\), which readily allows its uncertainty to be
quantified.

Spatial deformation models suffer the intuitively undesirable flaw of
allowing \(D\)-spaces that `fold', i.e., non-bijective mappings
\(\boldsymbol{g}\) or, more conceptually, mappings such that for every
\(\boldsymbol{x}\) in \(G\)-space there is not a single
\(\boldsymbol{x}^*\) in \(D\)-space. \cite{damian2001} hinder folding by
considering the `bending energy' of \(D\)-spaces, and increasingly
penalize spaces that require less energy to bend. Alternatively,
\cite{schmidt2003} propose to represent \(\boldsymbol{g}\) as a
multivariate GP, and state that `the GP formulation for
{[}\(\boldsymbol{g}\){]} tends to eliminate the kind of non-injective
mappings that were noted by \cite{sampson1992}'.

Various approaches have explicitly addressed avoiding mappings that
fold. \cite{iovleff2004} use a Delaunay triangulation of \(G\)-space to
identify and eliminate mappings that give folds in \(D\)-space.
\cite{perrin-mon} derive conditions on deformations based on radial
basis functions that avoid folds. Nonstationarity in dependence is also
considered when emulating computer models and referred to as input
warping (IW): the computer model's inputs are transformed to a scale on
which dependence is stationary; see, e.g., \cite{snelson2004}.
\cite{zammit2022} propose deep compositional spatial models for
representing \(\boldsymbol{g}\) in which the compositional formulation
can ensure bijectivity. \cite{zammit2022} propose to represent the
compositions through IW GPs and deep stochastic processes (DSPs), both
of which are based on basis representations with weights and unknown
basis function parameters. The IW GPs have unknown weights, which are
estimated by maximum likelihood, whereas the DSPs have random weights,
which are assumed to be of log-Gaussian form and estimated by
variational Bayes. \cite{zammit2022} then propose three approaches to
warping: axial warping units, with positive weights and monotonic basis
functions; radial basis functions, employing the constraints of
\cite{perrin-mon}; and Möbius transformation units, which make analogy
between mapping from \(\mathbb{C}\) to itself with mapping from
\(\mathbb{R}^2\) to itself. Each can be used with IW GPs and DSPs and
ensure bijectivity by virtue of the compositional structure.
\cite{dias2020} represent \(D\)-space as the tensor product of B-splines
and derive constraints on the knots that, when imposed, give
bijectivity. For a wide and recent review of statistical modeling of
nonstationary covariance -- including deformation-based approaches --
see \cite{schmidt2020}.

Folding could be considered a consequence of a two-dimensional
\(D\)-space being insufficient to bring isotropy. \cite{schmidt2011} and
\cite{bornn2012} propose extending \(D\)-space to \(2 + r\) dimensions
for \(r \geq 1\) so that
\(\boldsymbol{g} : \mathbb{R}^{2} \mapsto \mathbb{R}^{2 + r}\).
\cite{bornn2012} refer to this approach as dimension expansion.
\cite{schmidt2011} place a GP prior on \(\boldsymbol{g}\), allow
covariates in \(v\) and then base \(v\) on Mahalanobis distance to
generalize the usual Euclidean distance. \cite{bornn2012} estimate the
latent dimensions in a two-stage procedure that finds interim values
using a least squares fit between empirical and model-based variograms,
which are then approximated using thin plate splines.

This article introduces the \pkg{deform} \proglang{R} package, which is
designed to fit nonstationary spatial GPs using either spatial
deformations or dimension expansions to represent \(\boldsymbol{g}\). In
particular, it facilitates the visualization of fitted models; for
example, we may want to inspect \(\boldsymbol{g}\) for any physical
interpretation.

Various \proglang{R} packages exist that fit geostatistical models. A
selection include packages \pkg{fields} \citep{fields}, \pkg{spatial}
\citep{MASS2002} and \pkg{LatticeKrig} \citep{LatticeKrig}, which can
perform kriging, and \pkg{geoR} \citep{geoR}, \pkg{gstat}
\citep{gstat1, gstat2}, \pkg{spmodel} \citep{spmodel} and
\pkg{RandomFields} \citep{Randomfields1, Randomfields2} give functions
for fitting stationary GPs. Extensions for fitting nonstationary GPs
include \pkg{FRK} \citep{FRK1, FRK2}, \pkg{tgp} \citep{tgp1, tgp2} and
\pkg{enviroStat} \citep{le2006}, which implements \cite{sampson1992}'s
method (but is no longer supported by CRAN). Alternatively, Gaussian
Markov random fields can be fit in \pkg{mgcv} with basis specification
\texttt{s(...,\ bs\ =\ "mrf")} (see \cite{wood-book}) and with
\pkg{R-INLA}, \citep[see also \url{http://www.r-inla.org}]{martins2013},
which uses the integrated nested Laplace approximation of
\cite{rue2009}. \proglang{R} packages for non-Gaussian geostatistical
data also exist, such as \pkg{SpatialExtremes} \citep{SpatialExtremes}
and \pkg{CARBayes} \citep{CARBayes}. The fitting of nonstationary GPs to
spatial data can be achieved through various software packages for DGPs,
such as \pkg{dgpsi} \citep{dgpsi1, dgpsi2} for \proglang{R}, and various
\proglang{python} options, such as \pkg{GPy} \citep{Gpy} and
\pkg{GPflow} \citep{GPflow}.

The next section of this article introduces low-rank representations for
\(\boldsymbol{g}\) based on thin plate \emph{regression} splines
\citep{wood-tprs}. These apply to both the spatial deformation approach
of \cite{sampson1992}, which is extended to ensure bijectivity, and to
the dimension expansion approach of \cite{bornn2012}. Section 3
introduces objective methods for inference for such models. Section 4
introduces the key functions in package \pkg{deform}. Section 5
demonstrates \pkg{deform}'s functions on the solar radiation data
originally used in \cite{sampson1992}. Section 6 covers the modeling of
censored data and then section 7 summarizes the work presented.

\section{Methodology} \label{meth}

In the following section, consider again \(Y_t(\boldsymbol{x})\), a
process at time \(t=1, \ldots, n\) and location
\(\boldsymbol{x} = (x_1, x_2) \in \mathcal{G}\). Specifically,
\pkg{deform} only considers the fitting of zero-mean GPs, so
\(\mu(\boldsymbol{x}) = 0\) in \eqref{gp}. Forms assumed for
\(v(\boldsymbol{g}(\boldsymbol{x}), \boldsymbol{g}(\boldsymbol{x}'))\)
in \pkg{deform} are given in \S\ref{deform:fct}. In the two-dimensional
case, \(x_1\) and \(x_2\) are longitude and latitude coordinates,
respectively. Then consider the transformation
\(\boldsymbol{x}^* =\boldsymbol{g}(\boldsymbol{x})\), where
\(\boldsymbol{x}\) exists in \(G\)-space and \(\boldsymbol{x}^*\) exists
in \(D\)-space. The methodology presented readily extends to
\(G\)-spaces defined over any number of dimensions, as in
\cite{bornn2012}. Independence over time will be assumed to focus on
spatial dependence.

\subsection{Spatial deformation} \label{S:meth:deform}

A spatial deformation is defined here as the transformation
\(\boldsymbol{x}^* =\boldsymbol{g}(\boldsymbol{x})\), given a mapping
\(\boldsymbol{g}\), where
\(\boldsymbol{g}:\mathbb{R}^2 \mapsto \mathbb{R}^2\),
\(\boldsymbol{x}^* = (x_1^*, x_2^*) = (g_1(\boldsymbol{x}), g_2(\boldsymbol{x}))\)
and \(g_d : \mathbb{R}^2 \mapsto \mathbb{R}\) for \(d = 1, 2\). For
spatial deformations, \cite{smith1996}'s parametrization is adopted, so
that for
\(\boldsymbol{x}^* = (x_1^*, x_2^*) = (g_1(\boldsymbol{x}), g_2(\boldsymbol{x}))\),
\begin{align}
\label{g1} g_1(\boldsymbol{x}) = \exp(\alpha_1) x_1 + \alpha_3 x_2 + \sum_{i = 1}^m \delta_{1i} \eta(\norm{\boldsymbol{x} - \boldsymbol{x}_i}),\\ 
\label{g2} g_2(\boldsymbol{x}) =  \alpha_3 x_1 + \exp(\alpha_2) x_2 + \sum_{i = 1}^m \delta_{2i} \eta(\norm{\boldsymbol{x} - \boldsymbol{x}_i}), 
\end{align} where \(\eta(l) = l^2 \log l\) with
\(\norm{\mathbf{x} - \mathbf{x}'} = \sqrt{(x_1 - x_1')^2 + (x_2 - x_2')^2}\)
and the above equations are subject to the constraints
\(\mathbf{T}^\text{T} \boldsymbol{\delta}_1 = \mathbf{T}^\text{T} \boldsymbol{\delta}_2 = \mathbf{0}_{3 \times 1}\),
where \(m \times 3\) matrix \(\mathbf{T}\) has \(i\)th row
\((1, x_{i1}, x_{i2})\), for \(i = 1, \ldots, m\),
\(\boldsymbol{\delta}_1 = (\delta_{11}, \ldots, \delta_{1m})^\text{T}\),
\(\boldsymbol{\delta}_2 = (\delta_{21}, \ldots, \delta_{2m})^\text{T}\)
and \(\mathbf{0}_{n \times m}\) denotes a \(n \times m\) matrix
comprizing only zeros. These constraints avoid over-parametrization,
while the coefficients of \(x_1\) and \(x_2\) in equations \eqref{g1}
and \eqref{g2} avoid rotational invariance. Following \cite{wood-tprs},
let matrix \(\mathbf{E}\) have \((i, j)\)th element
\(E_{ij} = \eta(\norm{\boldsymbol{x}_i - \boldsymbol{x}_j})\) and
eigen-decomposition
\(\mathbf{E} = \mathbf{U} \boldsymbol{\Lambda} \mathbf{U}^\text{T}\).
Then let
\(\mathbf{E}_k = \mathbf{U}_k \boldsymbol{\Lambda}_k \mathbf{U}_k^\text{T}\),
where \(\mathbf{U}_k\) denotes the first \(k\) columns of \(\mathbf{U}\)
and \(\boldsymbol{\Lambda}_k\) denotes the upper \(k \times k\) block of
\(\boldsymbol{\Lambda}\), with
\(\boldsymbol{\Lambda} = \text{diag}(e_1, \ldots, e_m)\) comprizing the
\(m\) eigenvalues of \(\boldsymbol{\Lambda}\) arranged as
\(e_1 \geq \ldots \geq e_m\). Attention can then be restricted to the
space spanned by \(\mathbf{U}_k\), i.e., to
\(\boldsymbol{\delta}_1^{(k_1)}\) such that
\(\boldsymbol{\delta}_1 = \mathbf{U}_{k_1} \boldsymbol{\delta}_1^{(k_1)}\)
and \(\boldsymbol{\delta}_2^{(k_2)}\) such that
\(\boldsymbol{\delta}_2 = \mathbf{U}_{k_2} \boldsymbol{\delta}_2^{(k_2)}\),
where \(k_1\) and \(k_2\) are the finite ranks chosen to represent
\(g_1\) and \(g_2\), respectively (although often \(k_1 = k_2\) will be
used in practice). An unconstrained optimization problem can be formed
from the constraints
\(\mathbf{T}^\text{T} \boldsymbol{\delta}_1 = \mathbf{T}^\text{T} \boldsymbol{\delta}_2 = \mathbf{0}_{3 \times 1}\)
for the rank-\(k\) case by choosing some matrix \(\mathbf{Z}_k\) such
that
\(\mathbf{T}^\text{T} \mathbf{U}_k \mathbf{Z}_k = \mathbf{0}_{k \times (k - 3)}\).
This can be achieved via the QR-decomposition,
\(\mathbf{T}^\text{T} \mathbf{U}_k = \mathbf{QR}\), say, taking
\(\mathbf{Z}_k\) as the final \(k - 3\) columns of \(\mathbf{Q}\). For
the unconstrained problem, \(\tilde{\boldsymbol{\delta}}_1\) and
\(\tilde{\boldsymbol{\delta}}_2\) satisfying
\(\boldsymbol{\delta}_1^{(k_1)} = \mathbf{Z}_{k_1} \tilde{\boldsymbol{\delta}}_1\)
and
\(\boldsymbol{\delta}_2^{(k_2)} = \mathbf{Z}_{k_2} \tilde{\boldsymbol{\delta}}_2\)
can be used. The unknown parameters that define the deformation are
given by the \((k_1 + k_2 - 3)\)-vector
\(\boldsymbol{\beta} = (\alpha_1, \alpha_2, \alpha_3, \tilde{\boldsymbol{\delta}}_1^\text{T}, \tilde{\boldsymbol{\delta}}_2^\text{T})^\text{T}\).

Typically an additive wiggliness penalty is imposed on \(g_1\) and
\(g_2\) of the form
\(\lambda_1 \boldsymbol{\delta}_1^\text{T} \mathbf{E} \boldsymbol{\delta}_1 + \lambda_2 \boldsymbol{\delta}_2^\text{T} \mathbf{E} \boldsymbol{\delta}_2\).
This becomes
\(\lambda_1 (\boldsymbol{\delta}_1^{(k_1)})^\text{T} \boldsymbol{\Lambda}_{k_1} \boldsymbol{\delta}_1^{(k_1)} + \lambda_2 (\boldsymbol{\delta}_2^{(k_2)})^\text{T} \boldsymbol{\Lambda}_{k_2} \boldsymbol{\delta}_2^{(k_2)}\),
allowing for the finite-rank representations of \(g_1\) and \(g_2\), and
then
\(\lambda_1 (\tilde{\boldsymbol{\delta}}_1^{(k_1)})^\text{T} \mathbf{Z}_{k_1}^\text{T} \boldsymbol{\Lambda}_{k_1} \mathbf{Z}_{k_1} \tilde{\boldsymbol{\delta}}_1^{(k_1)} + \lambda_2 (\tilde{\boldsymbol{\delta}}_2^{(k_2)})^\text{T} \mathbf{Z}_{k_2}^\text{T} \boldsymbol{\Lambda}_{k_2} \mathbf{Z}_{k_2} \tilde{\boldsymbol{\delta}}_2^{(k_2)}\),
allowing for the unconstrained optimization problem. Let
\(\boldsymbol{S}_{\boldsymbol{\lambda}} = \lambda_1 \boldsymbol{S}_1 + \lambda_2 \boldsymbol{S}_2\),
where
\(\boldsymbol{S}_1 = \text{diag}(\mathbf{0}_{3 \times 3}, \allowbreak \mathbf{Z}_{k_1}^\text{T} \boldsymbol{\Lambda}_{k_1} \mathbf{Z}_{k_1}, \allowbreak \mathbf{0}_{(k_2 - 3) \times (k_2 - 3)})\)
and
\(\boldsymbol{S}_2 = \text{diag}(\mathbf{0}_{k_1 \times k_1}, \mathbf{Z}_{k_2}^\text{T} \boldsymbol{\Lambda}_{k_2} \mathbf{Z}_{k_2})\),
where \(\boldsymbol{\lambda} = (\lambda_1, \lambda_2)\). Given
\(\boldsymbol{\beta}\) above, the wiggliness penalty can be written
\(\boldsymbol{\beta}^\text{T} \boldsymbol{S}_{\boldsymbol{\lambda}} \boldsymbol{\beta}\),
which is of the form presented in \cite{wood-reml2} and hence used in
Section \ref{S:inf}. Fitting the spatial deformation model is achieved
by estimating \(\boldsymbol{\beta}\) subject to the penalty
\(\boldsymbol{\beta}^\text{T} \boldsymbol{S}_{\boldsymbol{\lambda}} \boldsymbol{\beta}\).

\subsection{Bijective spatial deformation} \label{S:meth:biject}

\cite{bornn2012} demonstrate how smoothing parameters associated with
thin plate splines, or equivalently that control the bending energy, may
be fixed to ensure bijectivity. This work aims to maintain objectivity
by allowing optimal estimation of smoothing parameters under the
condition of bijectivity. \cite{iovleff2004} ensure bijectivity by
representing \(G\)-space as a Delaunay triangulation, which, when
transformed to \(D\)-space, is bijective if none of the vertices lie
within any of the triangles. The approach of \cite{iovleff2004} applies
to any form for \(\boldsymbol{g}\), which is a criterion that the
approach proposed here also satisfies. The approaches of
\cite{perrin-mon} and \cite{zammit2022} require specific---albeit
seemingly rather flexible---forms for \(\boldsymbol{g}\).

\begin{CodeChunk}
\begin{figure}

{\centering \includegraphics{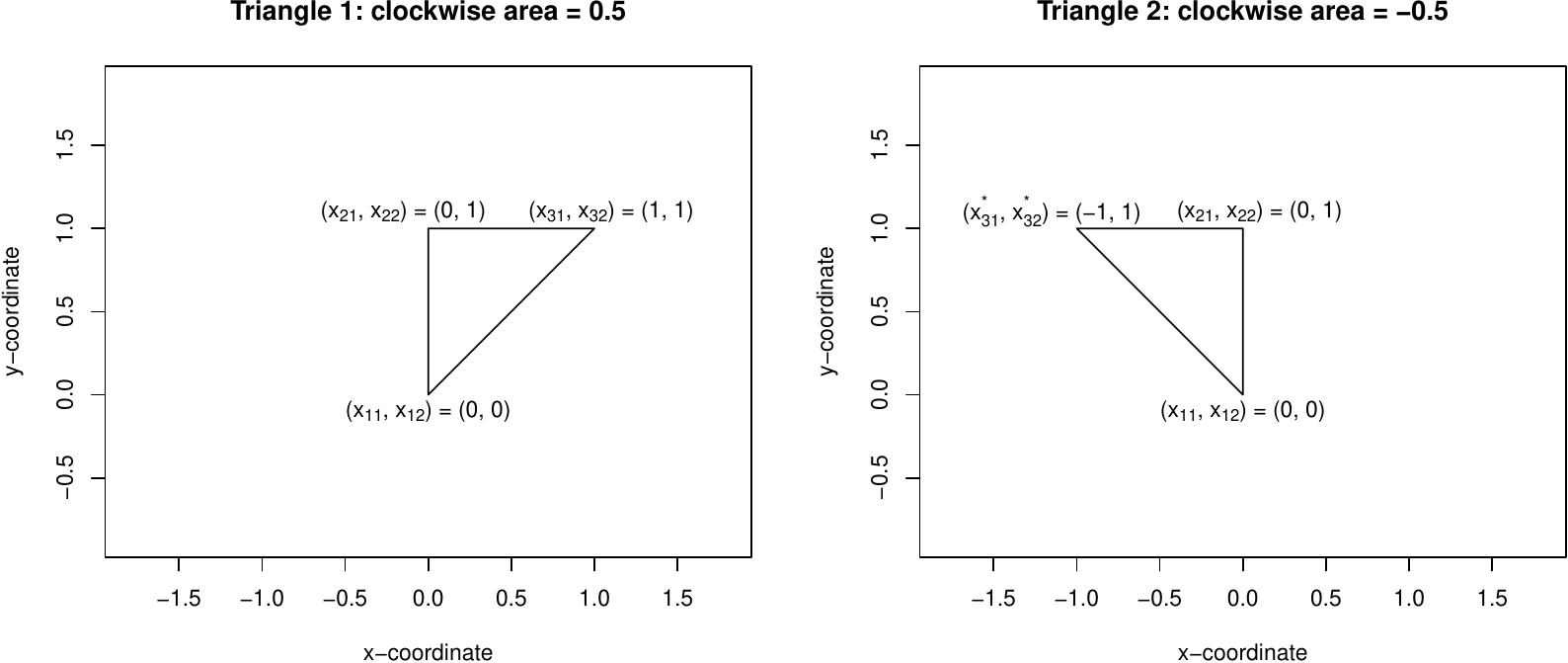} 

}

\caption{\label{clockwise}Clockwise areas of triangles. Left: A triangle represented as clockwise points $(x_{11}, x_{12})$, $(x_{21}, x_{22})$, $(x_{31}, x_{32})$ with clockwise area 0.5. Right: A transformation to the left-hand triangle, in which $(x_{31}, x_{32}) \mapsto (x_{31}^*, x_{32}^*)$, giving a negative clockwise area of -0.5, based on the clockwise ordering in the left-hand triangle. Such negative areas are used to identify grids that have folded.}\label{fig:clckchunk}
\end{figure}
\end{CodeChunk}

\begin{figure}[h!]
\begin{center}
\includegraphics[width=.8\textwidth]{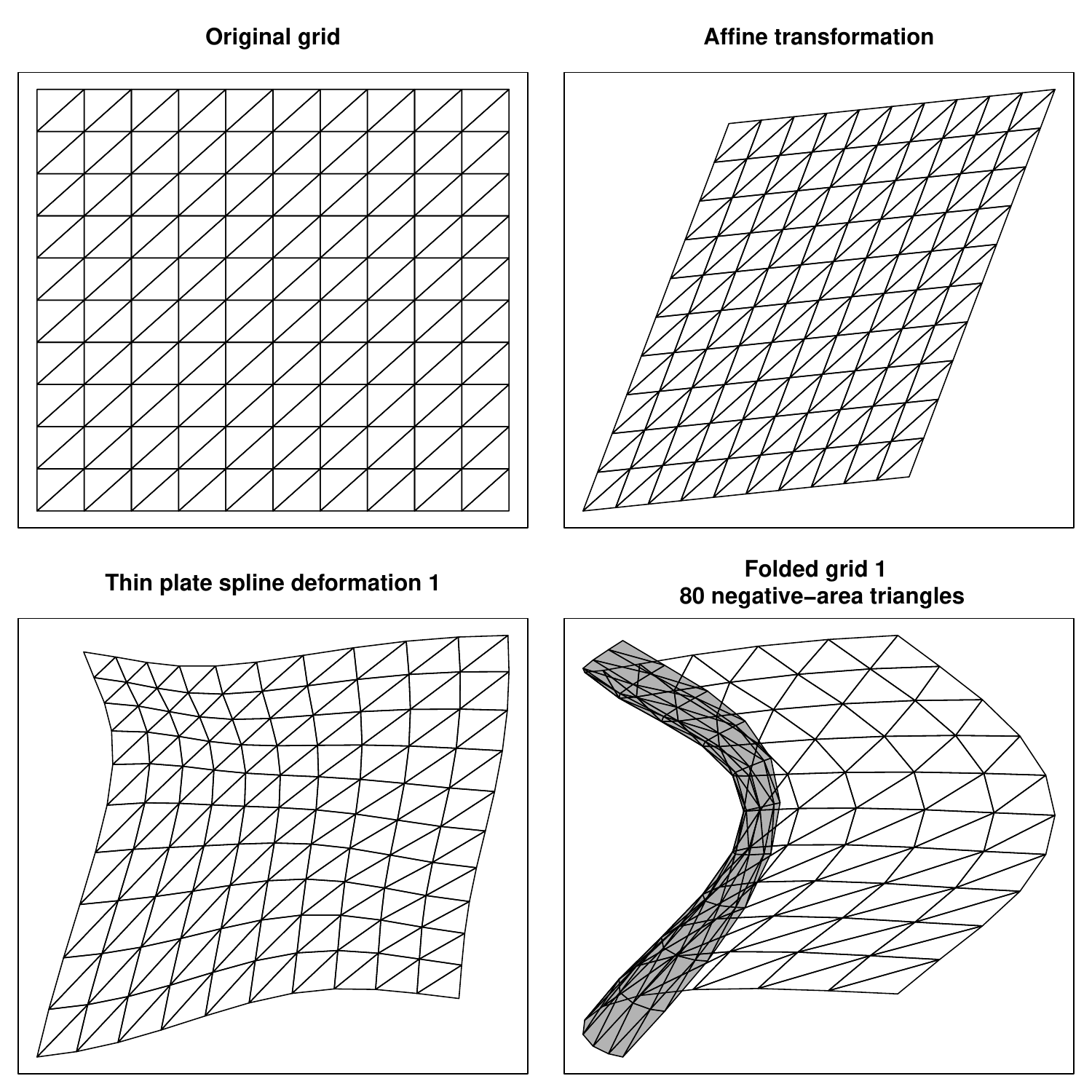}
\end{center}
\caption{\label{fold}Representations of spatial deformations using triangular tilings. Row 1, column 1: The original triangular tiling of the domain-spanning grid for calculating clockwise triangle areas. Row 1, column 2: An example of an affine transform of the form $\boldsymbol{x} \mapsto {\bf M} \boldsymbol{x}$, for a $2 \times 2$ matrix ${\bf M}$ and $\boldsymbol{x} = (x_1, x_2)'$. Row 2: Examples of deformations achieved by $\boldsymbol{x} \mapsto \boldsymbol{g}(\boldsymbol{x})$, where $\boldsymbol{g}$ comprises two thin plate regression splines, demonstrating fold-free (column 1) and folded (column 2) deformations.}
\end{figure}

A related approach to \cite{iovleff2004} is proposed here in which
\(\mathcal{G}\), the domain of interest, is represented as a triangular
tiling. The clockwise area of each triangle is computed, which, based on
Figure \ref{clockwise}, is given by
\((x_{21} x_{12} + x_{31} x_{22} + x_{11} x_{32} - x_{11} x_{22} - x_{21} x_{32} - x_{31} x_{12})/2\),
where \((x_{i1}, x_{i2})\), \(i=1, 2, 3\), are vertices of a triangle
defined in clockwise order. Subject to the triangular tiling's finite
representation of \(\mathcal{G}\), a change in ordering can be used to
identify non-bijective \(\boldsymbol{g}\), which is equivalent to
\(\boldsymbol{g}\) turning a triangle's clockwise area negative. This is
illustrated in Figure \ref{clockwise} in which the left-hand triangle
has clockwise area 0.5 whereas the right-hand triangle has clockwise
area \(-0.5\). A space represented by a triangular tiling (see Figure
\ref{fold}, row 1, column 1) with a mixture of positive and negative
areas must have folded (see Figure \ref{fold}, row 2, column 2); all
positive areas corresponds to a fold-free space; and all negative areas
corresponds to a fold-free space that has `flipped'. Flipped spaces can
be eliminated without loss of generality since equivalent distances for
such spaces can be achieved if the space is flipped back.

Consider the triangular tiling
\(\mathcal{G} = \cup_{l=1}^L \mathcal{W}_l\), where each
\(\mathcal{W}_l\), for \(l=1, \ldots, L\), is a triangle with clockwise
area \(A(\mathcal{W}_l)\). Imposing the penalty
\(\delta I\big[\min\{A(\mathcal{W}_1), \ldots, A(\mathcal{W}_L)\} < 0\big]\)
could be used to eliminate folds, for large \(\delta > 0\) and where
\(I(\,)\) is the indicator function. In practice this can work poorly
with the inference approach proposed in \S\ref{S:inf} as it can result
in parameter estimates at a non-differentiable part of parameter space.
Instead, a penalty of the form \begin{equation}
\label{foldpen}
\frac{\delta}{2} \big[A_-(\mathcal{W}_1 / \epsilon) + \ldots + A_-(\mathcal{W}_L / \epsilon)\big]^2
\end{equation} is chosen, where \(A_-(W) = \min\{A(W), 0\}\) and
\(\epsilon > 0\), which increasingly penalizes deformations with more
folding and is twice-differentiable, which aids inference. User-friendly
choice of \(\delta\) and \(\epsilon\) is discussed in
\S\ref{S:solar:biject}.

\subsection{Dimension expansion} \label{S:meth:expand}

Dimension expansions have
\(\boldsymbol{x}^* =\boldsymbol{g}(\boldsymbol{x})\) with
\(\boldsymbol{g}:\mathbb{R}^2 \mapsto \mathbb{R}^{2 + r}\) where
\(\boldsymbol{x}^* = (g_1(\boldsymbol{x}), \ldots, g_{2 + r}(\boldsymbol{x}))\),
with \(g_d : \mathbb{R}^2 \mapsto \mathbb{R}\) for
\(d = 1, \ldots, r + 2\). The parametrization for dimension expansions
is a slight extension to that of \cite{bornn2012}, so that
\(\boldsymbol{x}^* = (g_1(\boldsymbol{x}), g_2(\boldsymbol{x}), \ldots, g_{r + 2}(\boldsymbol{x}))\),
where \begin{align*}
g_1(\boldsymbol{x}) &= \exp(\alpha_1) x_1 + \alpha_3 x_2,\\
g_2(\boldsymbol{x}) &=  \alpha_3 x_1 + \exp(\alpha_2) x_2,\\
g_{d + 2}(\boldsymbol{x}) &= \sum_{i = 1}^m \delta_{di} \eta(\norm{\boldsymbol{x} - \boldsymbol{x}_i}),
\end{align*} for \(d = 1, \ldots, r\), with \(\eta\) as in
\S\ref{S:meth:deform}. As in \S\ref{S:meth:deform}, matrix
\(\mathbf{E}\) with \((i, j)\)th element
\(E_{ij} = \eta(\norm{\boldsymbol{x}_i - \boldsymbol{x}_j})\) can be
formed with eigen-decomposition
\(\mathbf{E} = \mathbf{U}^\text{T} \boldsymbol{\Lambda} \mathbf{U}\),
which can be used to give a rank-\(k_d\) approximation to
\(g_{d + 2}(\boldsymbol{x})\). For dimension expansions, the usual
affine transformation of the thin plate spline is omitted from
\(g_{d + 2}\) here, for \(d = 1, \ldots, r\); instead a scale and
rotation transformation is captured with \(g_1\) and \(g_2\). As a
result, parameters of \(g_{d + 2}\) for \(d = 1, \ldots, r\) are
\(\tilde{\boldsymbol{\delta}}_d\) satisfying
\(\boldsymbol{\delta}_d = \mathbf{U}_{k_d} \tilde{\boldsymbol{\delta}}\),
with \(\mathbf{U}_{k_d}\) the first \(k_d\) columns of \(\mathbf{U}\)
and where
\(\boldsymbol{\delta}_d = (\delta_{d1}, \ldots, \delta_{dm})^\text{T}\).
The unknown parameters that define the dimension expansion are given by
the \((3 + K_r)\)-vector
\(\boldsymbol{\beta} = (\alpha_1, \alpha_2, \alpha_3, \tilde{\boldsymbol{\delta}}_1^\text{T}, \ldots, \tilde{\boldsymbol{\delta}}_r^\text{T})^\text{T}\),
where \(K_d = \sum_{j = 1}^d k_j\). A wiggliness penalty can be imposed
on each \(g_{d + 2}\) of the form
\(\lambda_d \tilde{\boldsymbol{\delta}}_d^\text{T} \boldsymbol{\Lambda}_{k_d} \tilde{\boldsymbol{\delta}}_d\)
such that \(\lambda_d \to \infty\) gives \(g_d(\boldsymbol{x}) \to 0\),
for \(d = 1, \ldots, r\); see \cite{marra2011}. Large values of
\(\lambda_d\) can be used to identify and then eliminate redundant
dimensions. Penalties can be combined as
\(\boldsymbol{\beta}^\text{T} \boldsymbol{S}_{\boldsymbol{\lambda}} \boldsymbol{\beta}\),
where
\(\boldsymbol{S}_{\boldsymbol{\lambda}} = \lambda_1 \boldsymbol{S}_1 + \ldots + \lambda_r \boldsymbol{S}_r\),
where \(\boldsymbol{\lambda} = (\lambda_1, \ldots, \lambda_r)\) and
\(\boldsymbol{S}_d\) is a \((3 + K_r) \times (3 + K_r)\) matrix
\(\boldsymbol{S}_d\) comprizing entirely zeros except for
\((e_1, \ldots e_{k_d})\) on its leading diagonal at positions
\(S_{4 + K_{d - 1}, 4 + K_{d - 1}}, \ldots, S_{3 + K_d, 3 + K_d}\),
respectively, with \(K_0 \equiv 0\).

\subsection{Notes on finite-rank basis representations}

\cite{sampson1992}, \cite{schmidt2003} and \cite{bornn2012} have
previously used thin plate splines to define locations in \(D\)-space.
This work proposes to use thin plate \emph{regression} splines, which
are a finite-rank approximation based on a truncated eigen-decomposition
of corresponding thin plate splines, with various optimality criteria:
see \cite{wood-tprs}. Instead of thin plate splines, the tensor product
of two one-dimensional splines could be used to form the two-dimensional
transformation from \(G\)- to \(D\)-space. More generally, the proposed
framework allows any \(g(\boldsymbol{x})\) function to be treated as a
smooth, and therefore can be characterized similarly to smooths in a
generalized additive models; see, e.g., \cite{wood-book}. Such
specifications would need careful consideration in order to avoid
over-parametrization and rotational invariance.

\section{Inference} \label{S:inf}

\subsection{Data and notation} \label{S:inf:data}

Estimation of the spatial deformation or dimension expansion models will
be presented for a fixed set of locations,
\(\mathcal{X} = \{\boldsymbol{x}_1, \ldots, \boldsymbol{x}_m\}\), and a
fixed set of time points, \(\mathcal{T} = \{1, \ldots, n\}\).
Corresponding data are
\(\boldsymbol{y} = (\boldsymbol{y}_1, \ldots, \boldsymbol{y}_n)\), with
\(\boldsymbol{y}_t\) for \(t = 1, \ldots, n\) as defined in
\S\ref{S:intro}. Note that \pkg{deform} absorbs any parameters required
by the covariance function \(v\) in relation \eqref{gp} into
\(\boldsymbol{\beta}\) and pads
\(\boldsymbol{S}_{\boldsymbol{\lambda}}\) with zeros accordingly.
Fitting models in \pkg{deform} therefore involves estimating
coefficients \(\boldsymbol{\beta}\), which determine the \(G\)- to
\(D\)-space mapping \(\boldsymbol{g}\), and smoothing parameters,
\(\boldsymbol{\lambda}\). Finally let \(\ell(\boldsymbol{\beta})\)
denote the log-likelihood of a zero-mean GP, as defined in equation
\eqref{gp}.

\subsection{Restricted maximum likelihood} \label{S:inf:pen}

For \(\boldsymbol{\beta}\) given \(\boldsymbol{\lambda}\), let
\(\ell_p(\boldsymbol{\beta}, \boldsymbol{\lambda})\) denote the
penalized log-likelihood given by
\begin{equation*} \label{penalized} \ell_p(\boldsymbol{\beta}, \boldsymbol{\lambda}) = \ell(\boldsymbol{\beta}) - \frac{1}{2}\boldsymbol{\beta}^T \boldsymbol{S}_{\boldsymbol{\lambda}} \boldsymbol{\beta}.\end{equation*}
In the case of bijective spatial deformations, this may be further
penalized with the penalty of equation \eqref{foldpen} to restrict
smoothing parameters \(\boldsymbol{\lambda}\) to those that lead to
bijective \(\boldsymbol{g}\).

Smoothing parameters, \(\boldsymbol{\lambda}\), are estimated
objectively using restricted maximum likelihood (REML), as outlined in
\cite{wood-reml} and \cite{wood-reml2}. This results from recognizing
that the penalized likelihood's penalty is proportional to the exponent
of a MVN(\({\bf 0}, \boldsymbol{S}_{\boldsymbol{\lambda}}^-)\)
distribution, where \(\boldsymbol{S}_{\boldsymbol{\lambda}}^-\) denotes
the generalized inverse of \(\boldsymbol{S}_{\boldsymbol{\lambda}}\).
Then \(\boldsymbol{\beta}\) is treated as a vector of random effects and
integrated out by Laplace's method. The resulting restricted
log-likelihood takes the form
\begin{equation*} \label{ell} \ell(\boldsymbol{\lambda}) = \ell_p(\boldsymbol{\beta}_{\boldsymbol{\lambda}}, \boldsymbol{\lambda}) + \frac{1}{2} \log |\boldsymbol{S}_{\boldsymbol{\lambda}}|_+ - \frac{1}{2} \log|{\bf H}(\hat{\boldsymbol{\beta}}_{\boldsymbol{\lambda}})| + \frac{M_p}{2} \log(2 \pi),\end{equation*}
where \(|\boldsymbol{S}_{\boldsymbol{\lambda}}|_+\) denotes the product
of positive eigenvalues of \(\boldsymbol{S}_{\boldsymbol{\lambda}}\),
\(\hat{\boldsymbol{\beta}}_{\boldsymbol{\lambda}}\) maximises
\(\ell_p(\boldsymbol{\beta}, \boldsymbol{\lambda})\) w.r.t.
\(\boldsymbol{\beta}\),
\({\bf H}(\hat{\boldsymbol{\beta}}_{\boldsymbol{\lambda}})\) is the
negative Hessian of \(\ell_p(\boldsymbol{\beta}, \boldsymbol{\lambda})\)
evaluated at \(\hat{\boldsymbol{\beta}}_{\boldsymbol{\lambda}}\) and
\(M_p\) is number of zero eigenvalues in
\(\boldsymbol{S}_{\boldsymbol{\lambda}}\).

Let \(\hat{\boldsymbol{\lambda}}\) denote the value of
\(\boldsymbol{\lambda}\) that maximises \(\ell(\boldsymbol{\lambda})\)
w.r.t. \(\boldsymbol{\lambda}\). Obtaining
\(\hat{\boldsymbol{\lambda}}\) is an iterative procedure in which each
evaluation of \(\ell(\boldsymbol{\lambda})\) involves obtaining
\(\hat{\boldsymbol{\beta}}_{\boldsymbol{\lambda}}\), which is achieved
in \pkg{deform} using Newton's method with first and second derivatives
of \(\ell_p(\boldsymbol{\beta}, \boldsymbol{\lambda})\) w.r.t.
\(\hat{\boldsymbol{\beta}}_{\boldsymbol{\lambda}}\) calculated
analytically. Higher-order derivatives in \pkg{deform} are approximated
by finite differencing due to their analytical complexity, so that
quasi-Newton methods are used to estimate \(\boldsymbol{\lambda}\).

Uncertainty in \(\boldsymbol{g}\) is quantified in \cite{damian2001}
using samples of parameters from their posterior distributions. Here,
however, its uncertainty is readily calculated once
\(\hat{\boldsymbol{\lambda}}\) has been obtained by assuming that the
sampling distribution of
\(\hat{\boldsymbol{\beta}}_{\hat{\boldsymbol{\lambda}}}\) is
\(MVN(\hat{\boldsymbol{\beta}}_{\hat{\boldsymbol{\lambda}}}, {\bf H}(\hat{\boldsymbol{\beta}}_{\hat{\boldsymbol{\lambda}}})^{-1})\);
see \S\ref{S:deform:predict}.

\section{Functions and model specifications} \label{deform:fct}

The package \pkg{deform} mainly relies on the functions \code{aniso()},
\code{deform()} and \code{expand()}. These all fit zero-mean GPs.
Specifically, \code{aniso()} fits a conventional anisotropic model,
i.e.~where \((x_1^*, x_2^*) = (\alpha_1 x_1, \alpha_2 x_2)\), with
\(\alpha_1, \alpha_2 > 0\). Then \code{deform()} and \code{expand()} fit
the spatial deformation and dimension expansion models of
\S\ref{S:meth:deform} and \S\ref{S:meth:expand}, respectively.

Their core usage is given by

\begin{Code}
  aniso(x, z, n)
  deform(x, z, n, k)
  expand(x, z, n, k)
\end{Code}

where \code{x} is an \(m \times 2\) matrix of coordinates, with each row
giving longitude and then latitude, \code{z} is an \(m \times m\)
empirically-calculated variance-covariance matrix, \code{n} \(= n\) is
the number of realizations from which \code{z} has been calculated and
\code{k} is a vector that specifies the ranks of latent dimensions.
Alternatively, \code{x} can be supplied as a \code{list} with elements
\code{x}, \code{z} and \code{n}, as described above.

Each model fitting function by default fits the powered exponential
covariance function, so that
\(v(\boldsymbol{g}(\boldsymbol{x}), \boldsymbol{g}(\boldsymbol{x}')) = \sigma^2 \rho(||\boldsymbol{g}(\boldsymbol{x}) - \boldsymbol{g}(\boldsymbol{x}')||)\),
where \[
\rho(l) = 
\left\{\begin{array}{ll}
(1 - \kappa) \exp(-l^\gamma) & \text{if} ~~ l \neq 0,\\
1 & \text{if} ~~ l = 0,
\end{array}\right.
\] for \(0 \leq \kappa < 1\) and \(0 < \gamma \leq 2\), and
\(\norm{\cdot}\) denotes Euclidean distance. The powered exponential
form is chosen for its greater flexibility over the exponential form and
analytical tractability over the Matérn form. Specifying
\code{correlation = TRUE} fixes \(\sigma^2 = 1\) and specifying
\code{cosine = TRUE} takes
\(v(\boldsymbol{g}(\boldsymbol{x}), \boldsymbol{g}(\boldsymbol{x}')) = \sigma^2 \rho(||\boldsymbol{g}(\boldsymbol{x}) - \boldsymbol{g}(\boldsymbol{x}')||) \cos(||\boldsymbol{g}(\boldsymbol{x}) - \boldsymbol{g}(\boldsymbol{x}')|| / \phi)\),
for \(\phi > 0\).

The default values in \code{deform()} are \code{k = c(10, 10)}, so that
\(g_1(\boldsymbol{x})\) and \(g_2(\boldsymbol{x})\) are both represented
as rank-10 thin plate regression splines. These ranks can be changed.
The default value in \code{expand()} is \code{k = 10}, so that \(r = 1\)
and \(g_3(\boldsymbol{x})\) is also represented as a rank-10 thin plate
regression spline. Various illustrations below show \code{aniso()},
\code{deform()} and \code{expand()} in action.

\section{Illustrations}

\subsection{Model fitting and visualization}

This section demonstrates the methods introduced in \S\ref{meth} and
\S\ref{S:inf} on solar radiation data for British Columbia. These data
were used in \cite{sampson1992}'s original paper on spatial
deformations, and originated from \cite{hay1984}. The radiation data
serve as proof-of-concept data, due to their popularization in
subsequent related works, such as \cite{schmidt2003} and
\cite{bornn2012}. Similarly to \cite{schmidt2003}, here the \(n=732\)
spring-summer measurements (22 March 1980 -- 20 September 1983,
excluding 21 September to 21 March each year) on solar radiation from 12
monitoring stations are studied. These data are available in
\pkg{deform} as dataset \code{solar}. The following accesses the data

\begin{CodeChunk}
\begin{CodeInput}
R> library(deform)
R> data(solar)
\end{CodeInput}
\end{CodeChunk}

and then the following shows the locations of the 12 monitoring
stations.

\begin{CodeChunk}
\begin{CodeInput}
R> library(mapdata)
R> maps::map('worldHires', 
+           xlim = range(pretty(solar$x[, 1])), 
+           ylim = range(pretty(solar$x[, 2])))
R> box()
R> maps::map.cities(canada.cities, minpop = 2e4)
R> points(solar$x, pch = 19)
\end{CodeInput}
\begin{figure}

{\centering \includegraphics[width=0.7\linewidth]{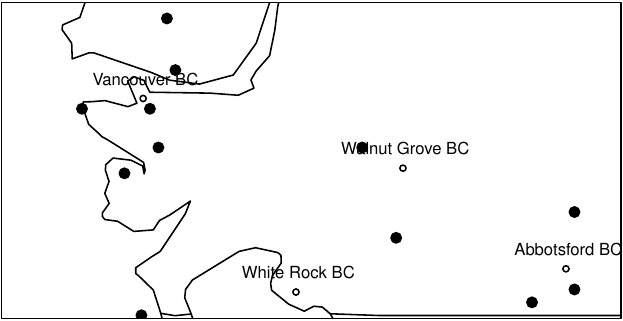} 

}

\caption[Locations of the 12 monitoring stations]{Locations of the 12 monitoring stations.}\label{fig:solar}
\end{figure}
\end{CodeChunk}

\subsubsection{Anisotropic modeling} \label{S:solar:aniso}

To start, we'll fit the conventional anisotropic model. This serves as a
benchmark against which subsequent nonstationary covariance models will
be compared. The dataset \code{solar} is a \code{list} in the form
described in \S\ref{deform:fct}, so the anisotropic model is fit simply
with the following.

\begin{CodeChunk}
\begin{CodeInput}
R> m0 <- aniso(solar)
\end{CodeInput}
\end{CodeChunk}

Once a model has been fit, \code{plot()} can be used to see a
representation of the model.

\begin{CodeChunk}
\begin{CodeInput}
R> plot(m0, asp = 1)
\end{CodeInput}
\begin{figure}

{\centering \includegraphics[width=0.7\linewidth]{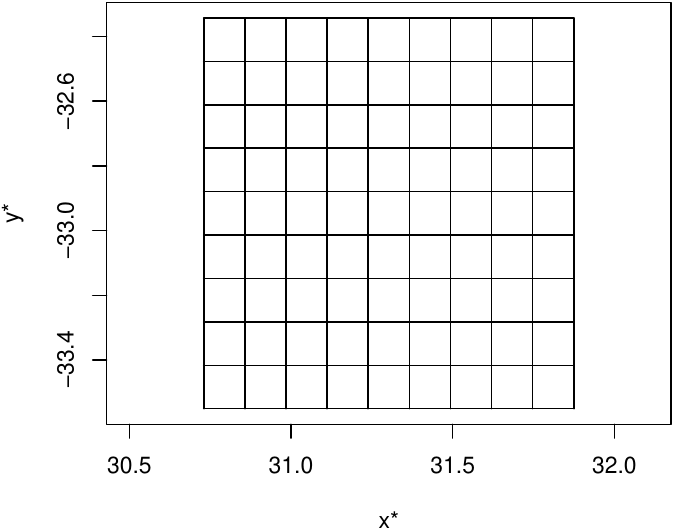} 

}

\caption[Anisotropic model representation]{Anisotropic model representation.}\label{fig:aniso2}
\end{figure}
\end{CodeChunk}

This takes a regular grid and then shows how this changes as a result of
the fitted model. In the case of the conventional anisotropic model,
this is rather trivial, as coordinates are just scaled.

\subsubsection{Spatial deformation} \label{S:solar:deform}

The following code then fits the default spatial deformation model.

\begin{CodeChunk}
\begin{CodeInput}
R> m1 <- deform(solar)
\end{CodeInput}
\end{CodeChunk}

We can supply \code{plot()} with longitude and latitude coordinates that
define the plotting grid through \code{xp} and \code{yp}, respectively.

\begin{CodeChunk}
\begin{CodeInput}
R> x_plot <- seq(-123.3, -122.25, by = .05)
R> y_plot <- seq(49, 49.4, by = .05)
R> plot(m1, xp = x_plot, yp = y_plot, asp = 1)
\end{CodeInput}
\begin{figure}

{\centering \includegraphics[width=0.7\linewidth]{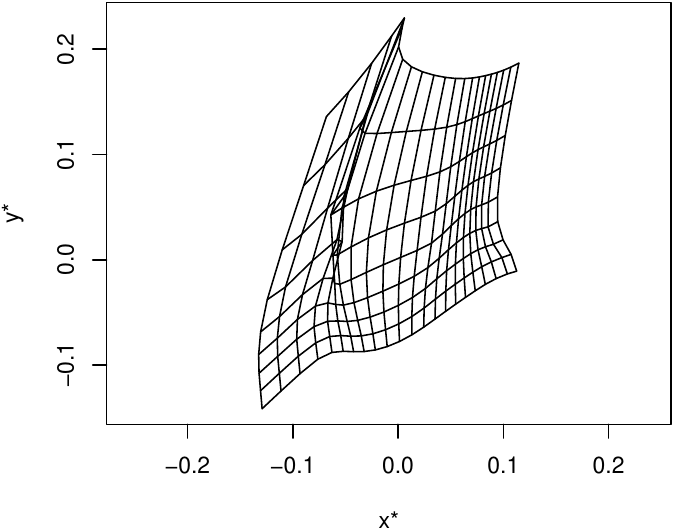} 

}

\caption[Deformation model representation]{Deformation model representation.}\label{fig:deform2}
\end{figure}
\end{CodeChunk}

The resulting representation of the fitted deformation shows a space
very different from that achieved by the simple scaling possible with
the conventional anisotropic model. Variograms can be used in
\pkg{deform} to compare models, which are covered in
\S\ref{S:solar:variogram}.

\subsubsection{Bijective spatial deformation} \label{S:solar:biject}

The representation of the deformation shown in Figure \ref{fig:deform2}
is seen to be non-bijective. We can ask \code{deform()} to numerically
eliminate bijective deformations by specifying \code{bijective = TRUE}.
This implements the numerical method described in \S\ref{S:meth:biject}.

\begin{CodeChunk}
\begin{CodeInput}
R> m3 <- deform(solar, bijective = TRUE)
R> plot(m3, xp = x_plot, yp = y_plot, asp = 1)
\end{CodeInput}
\begin{figure}

{\centering \includegraphics[width=0.7\linewidth]{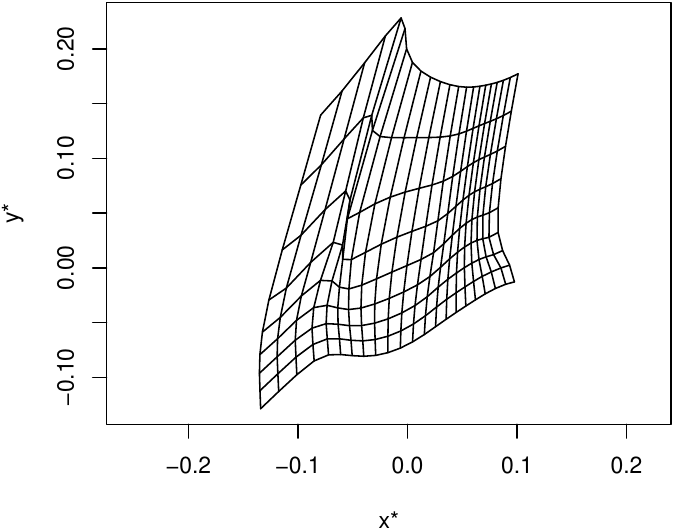} 

}

\caption{\label{fig:biject1}Bijective deformation model representation.}\label{fig:bijective1}
\end{figure}
\end{CodeChunk}

The argument \code{bijective.args} can be supplied to \code{deform()} as
a \code{list} to specify the parameters that control identifying
bijectivity, i.e.~\(\delta\) and \(\epsilon\) in equation
\eqref{foldpen}. Its default values are
\code{list(mult = 1e3, scl = 1, nx = 40, ny = 40)}, which uses a
triangular tiling of dimension \code{nx} \(\times\) \code{ny} to cover
the original data points, where \(\delta =\) \code{mult} and
\(\epsilon =\) \code{scl}
\(\times~ \hat \alpha_1 \hat \alpha_2 l_1 l_2\), with \(\hat \alpha_1\)
and \(\hat \alpha_2\) resulting from conventional anisotropic model
fits, and \(l_1\) and \(l_2\) are the resolutions of the heights and
widths, respectively, of the triangles in the tiling used to identify
bijectivity. Sensitivity to choice of \(\delta\) and \(\epsilon\) is
briefly studied in Appendix \ref{pensens}.

\subsubsection{One-dimensional expansion} \label{S:solar:expand1}

The following fits a one-dimensional expansion model, i.e.~with
\(r = 1\), based on the definition of \S\ref{S:meth:expand}.

\begin{CodeChunk}
\begin{CodeInput}
R> m2 <- expand(solar)
\end{CodeInput}
\end{CodeChunk}

Note that this model has one \emph{latent} dimension, although
\(\boldsymbol{g}(\boldsymbol{x}) = (g_1(\boldsymbol{x}), g_2(\boldsymbol{x}), g_3(\boldsymbol{x}))\),
and so technically uses three dimensions. The following plots the
model's three dimensions.

\begin{CodeChunk}
\begin{CodeInput}
R> par(mfrow = c(1, 3))
R> plot(m2)
\end{CodeInput}
\begin{figure}

{\centering \includegraphics{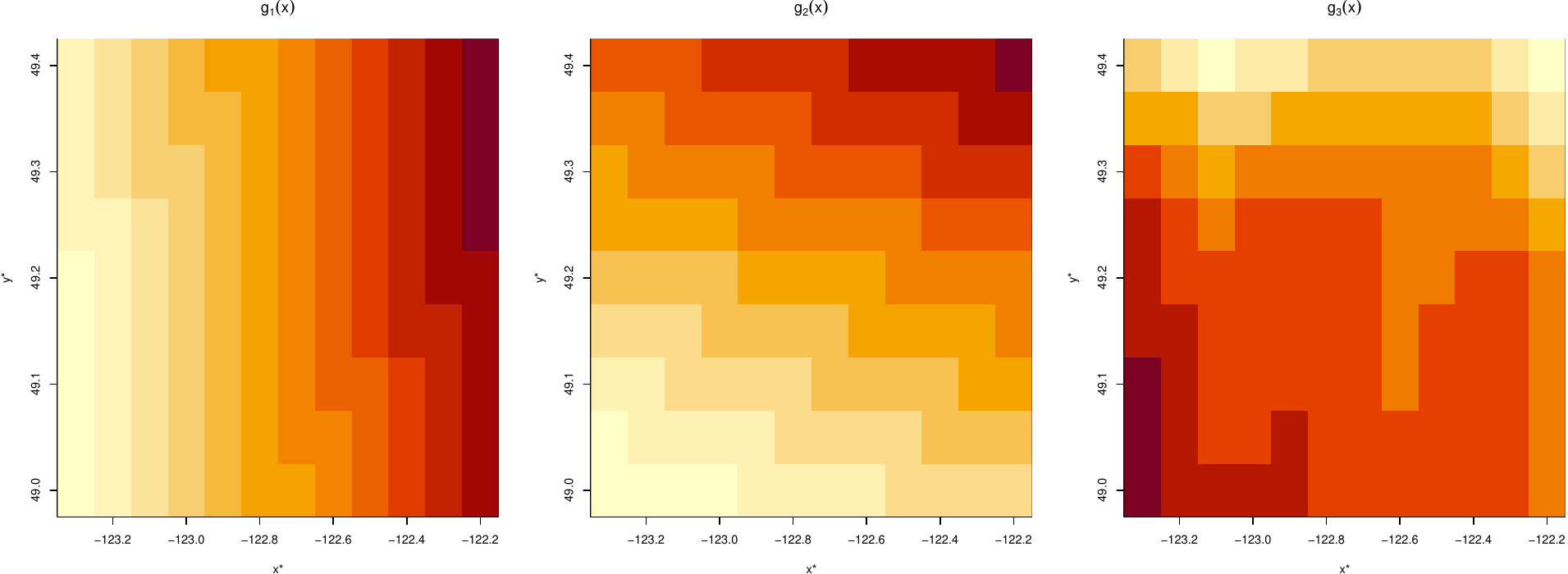} 

}

\caption[One-dimensional expansion model representation]{One-dimensional expansion model representation.}\label{fig:expand2}
\end{figure}
\end{CodeChunk}

\subsubsection{Two-dimensional expansion} \label{S:solar:expand2}

We can extend the number of latent dimensions by specifying a longer
vector for \code{k}. The following specifies two latent dimensions,
i.e.~\(r = 2\), each of which is represented as a rank-10 thin plate
regression spline, and then fits the model.

\begin{CodeChunk}
\begin{CodeInput}
R> m4 <- expand(solar, k = c(10, 10))
\end{CodeInput}
\end{CodeChunk}

When plotting such a model, we may only be interested in visualizing its
latent dimensions. The following starts plotting at the latent
dimensions, i.e.~omits plotting \(g_1(\boldsymbol{x})\) and
\(g_2(\boldsymbol{x})\). We can also use the \pkg{lattice} package to
produce plots (which gives colour scales), and we further specify that
all plots should be on one panel (\code{onepage = TRUE}).

\begin{CodeChunk}
\begin{CodeInput}
R> plot(m4, start = 3, graphics = 'lattice', onepage = TRUE)
\end{CodeInput}
\begin{figure}

{\centering \includegraphics{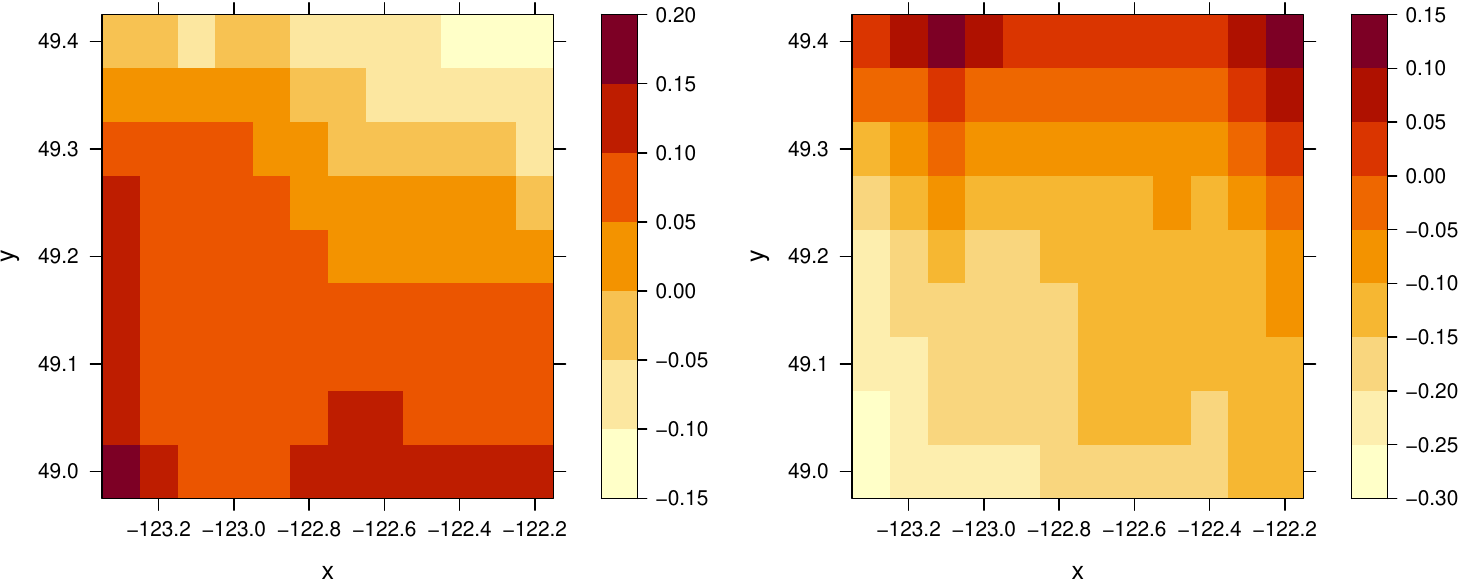} 

}

\caption[Two-dimensional expansion model representation]{Two-dimensional expansion model representation.}\label{fig:expand4}
\end{figure}
\end{CodeChunk}

\subsection{Variograms} \label{S:solar:variogram}

A simple way to assess the fit of a model is through a variogram,
comparing model-based estimates with empirical counterparts.
Specifically these plot semivariance against distance, where distance is
calculated in \(D\)-space. These can be achieved with function
\code{variogam()}, as shown below for the conventional anisotropic,
bijective deformation, and one- and two-dimensional expansion models.

\begin{CodeChunk}
\begin{CodeInput}
R> par(mfrow = c(2, 2))
R> variogram(m0); title('Conventional anisotropic model')
R> variogram(m3); title('Bijective deformation model')
R> variogram(m2); title('One-dimensional expansion model')
R> variogram(m4); title('Two-dimensional expansion model')
\end{CodeInput}
\begin{figure}

{\centering \includegraphics{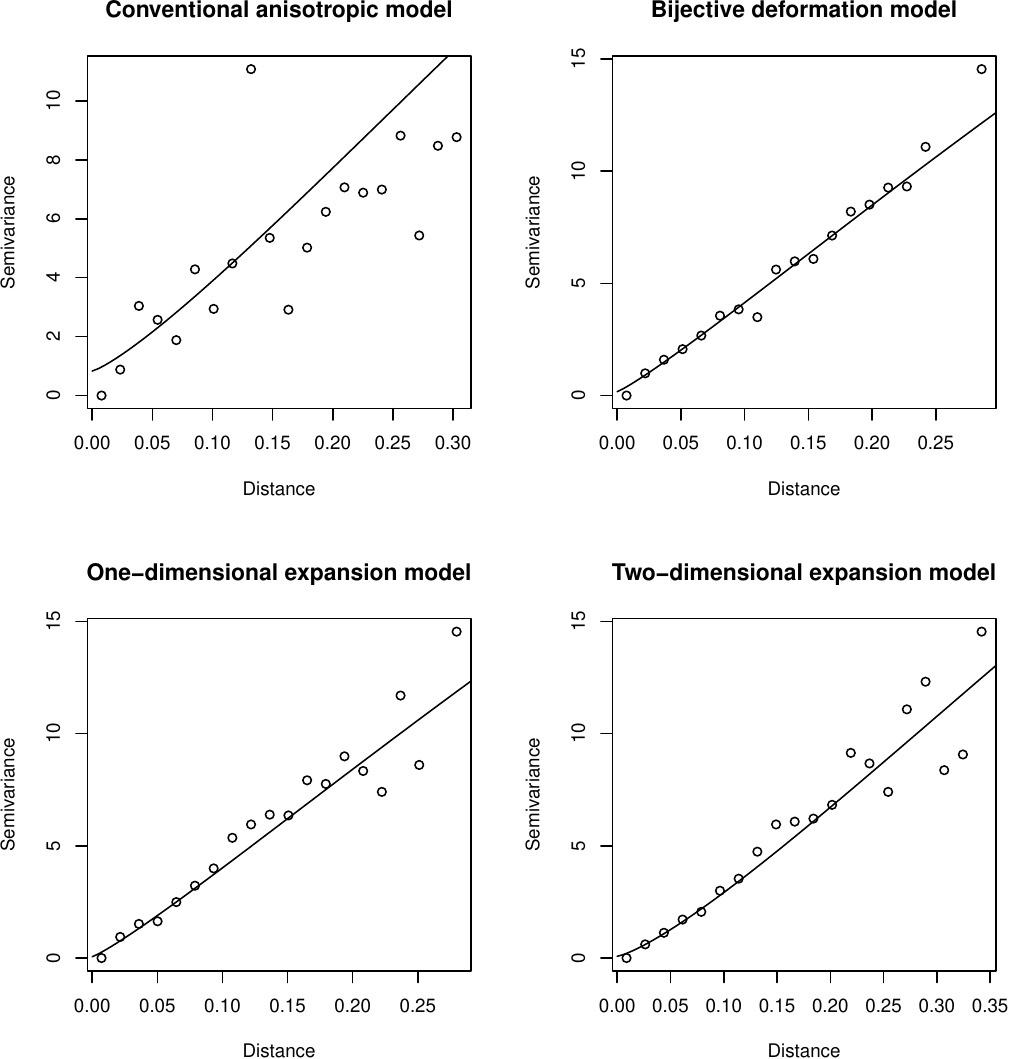} 

}

\caption[Variograms for conventional anisotropic, bijective deformation, and one- and two-dimensional expansion models]{Variograms for conventional anisotropic, bijective deformation, and one- and two-dimensional expansion models.}\label{fig:variogram1}
\end{figure}
\end{CodeChunk}

The variograms show that the bijective deformation model improves on the
conventional anisotropic model, and so too do the dimension expansion
models. However, there is little to distinguish between the one- and
two-dimensional expansions. Note that variograms for the non-bijective
and bijective deformation models are very similar, and hence only the
latter's variogram is shown.

\subsection{Prediction} \label{S:deform:predict}

Two forms of prediction are possible from objects of class \code{deform}
through \code{predict()}. The default, demonstrated here with the
bijective deformation model, is to give the coordinates in \(D\)-space
as a \code{data.frame}.

\begin{CodeChunk}
\begin{CodeInput}
R> predict(m3)
\end{CodeInput}
\begin{CodeOutput}
             [,1]         [,2]
 [1,] -0.01085683  0.202837264
 [2,] -0.05234367  0.066223514
 [3,] -0.05595379  0.038905276
 [4,] -0.08727547 -0.023210978
 [5,] -0.08045686 -0.082107777
 [6,]  0.02509023  0.022495256
 [7,]  0.07745855  0.022532967
 [8,]  0.09015940 -0.003449057
 [9,]  0.07610715 -0.014919774
[10,]  0.03105723 -0.014947634
[11,] -0.06635145 -0.002552110
[12,] -0.10746621 -0.005829157
\end{CodeOutput}
\end{CodeChunk}

The \code{predict()} argument \code{newdata} can be used to get
\(D\)-space coordinates for arbitrary \(G\)-space coordinates. If
argument \code{se.fit = TRUE} is given to \code{predict()}, then a
\code{list} is returned with elements \code{fitted}, which is as above,
and \code{se.fit}, a \code{data.fame} of pointwise standard errors for
estimated \(D\)-space coordinates. For example, the following gives the
predictions plotted in Figure \ref{fig:bijective1} together with
estimates of their standard error, with just the estimates corresponding
to the first six \(G\)-space coordinates shown below.

\begin{CodeChunk}
\begin{CodeInput}
R> grid_df <- as.matrix(expand.grid(x = x_plot, y = y_plot))
R> lapply(predict(m3, newdata = grid_df, se.fit = TRUE), head)
\end{CodeInput}
\begin{CodeOutput}
$fitted
            [,1]        [,2]
[1,] -0.13414235 -0.12866835
[2,] -0.11767010 -0.11295621
[3,] -0.10106640 -0.09763759
[4,] -0.08525571 -0.08502421
[5,] -0.07213916 -0.07950228
[6,] -0.06106455 -0.07919355

$se.fit
            [,1]        [,2]
[1,] 0.013533467 0.013215403
[2,] 0.012245433 0.011957164
[3,] 0.010986218 0.010745496
[4,] 0.009807857 0.009726404
[5,] 0.008749314 0.009148672
[6,] 0.007733122 0.008869357
\end{CodeOutput}
\end{CodeChunk}

Alternatively, \code{predict()} can be used to give the estimated
variance-covariance from a model.

\begin{CodeChunk}
\begin{CodeInput}
R> predict(m3, type = 'vcov')
\end{CodeInput}
\begin{CodeOutput}
          [,1]     [,2]     [,3]     [,4]     [,5]     [,6]     [,7]     [,8]
 [1,] 55.53827 49.54464 48.35936 45.40355 43.10049 47.75753 47.02726 45.78450
 [2,] 49.54464 55.53827 54.41427 51.57226 49.18786 51.87670 49.79774 48.85493
 [3,] 48.35936 54.41427 55.53827 52.70079 50.38424 52.14364 49.90845 49.13754
 [4,] 45.40355 51.57226 52.70079 55.53827 53.13184 50.47817 48.31829 47.98987
 [5,] 43.10049 49.18786 50.38424 53.13184 55.53827 49.29108 47.51718 47.58486
 [6,] 47.75753 51.87670 52.14364 50.47817 49.29108 55.53827 53.41883 52.68034
 [7,] 47.02726 49.79774 49.90845 48.31829 47.51718 53.41883 55.53827 54.36137
 [8,] 45.78450 48.85493 49.13754 47.98987 47.58486 52.68034 54.36137 55.53827
 [9,] 45.57995 49.14600 49.55185 48.63880 48.34416 52.96570 54.02394 54.77055
[10,] 46.12323 50.69191 51.30008 50.59472 50.09272 54.00642 53.11693 53.09347
[11,] 46.51092 52.67456 53.81226 54.34254 52.22437 51.62408 49.40505 48.94672
[12,] 45.77364 51.79984 52.75733 54.44953 52.21913 49.85905 47.61900 47.16249
          [,9]    [,10]    [,11]    [,12]
 [1,] 45.57995 46.12323 46.51092 45.77364
 [2,] 49.14600 50.69191 52.67456 51.79984
 [3,] 49.55185 51.30008 53.81226 52.75733
 [4,] 48.63880 50.59472 54.34254 54.44953
 [5,] 48.34416 50.09272 52.22437 52.21913
 [6,] 52.96570 54.00642 51.62408 49.85905
 [7,] 54.02394 53.11693 49.40505 47.61900
 [8,] 54.77055 53.09347 48.94672 47.16249
 [9,] 55.53827 53.71861 49.53505 47.76149
[10,] 53.71861 55.53827 51.47838 49.71658
[11,] 49.53505 51.47838 55.53827 53.87270
[12,] 47.76149 49.71658 53.87270 55.53827
\end{CodeOutput}
\end{CodeChunk}

\subsection{Simulation}

The \code{simulate()} function can be used to simulate realizations of
zero-mean GPs from a fitted object of class \code{deform}. Put simply,
realizations of nonstationary GPs in \(G\)-space are given by simulating
stationary GPs in \(D\)-space. The following gives one realization of a
nonstationary GP based on the transformation to \(D\)-space of the
coordinates of the 12 monitoring stations given by the bijective
deformation model, which was represented in Figure \ref{fig:biject1}.

\begin{CodeChunk}
\begin{CodeInput}
R> simulate(m3)
\end{CodeInput}
\begin{CodeOutput}
            [,1]
 [1,]  3.8218149
 [2,] -0.1519009
 [3,] -1.6879336
 [4,] -3.6426413
 [5,] -1.4220332
 [6,] -3.1855555
 [7,] -4.3454698
 [8,] -2.4847252
 [9,] -2.7502581
[10,] -2.3198710
[11,] -2.9696018
[12,] -3.2947852
\end{CodeOutput}
\end{CodeChunk}

The argument \code{nsim} specifies the number of realizations to
simulate, which defaults to one, and argument \code{newdata} can be used
similarly to \code{predict()} to specify the locations in \(G\)-space
for which simulations are sought. The following example shows three
realizations simulated for the grid used to represent fitted models
earlier.

\begin{CodeChunk}
\begin{CodeInput}
R> par(mfrow = c(1, 3))
R> sims <- simulate(m3, nsim = 3, newdata = grid_df)
R> for (i in 1:3) {
+   sim_mat <- matrix(sims[, i], length(x_plot))
+   image(x_plot, y_plot, sim_mat)
+ }
\end{CodeInput}
\begin{figure}

{\centering \includegraphics{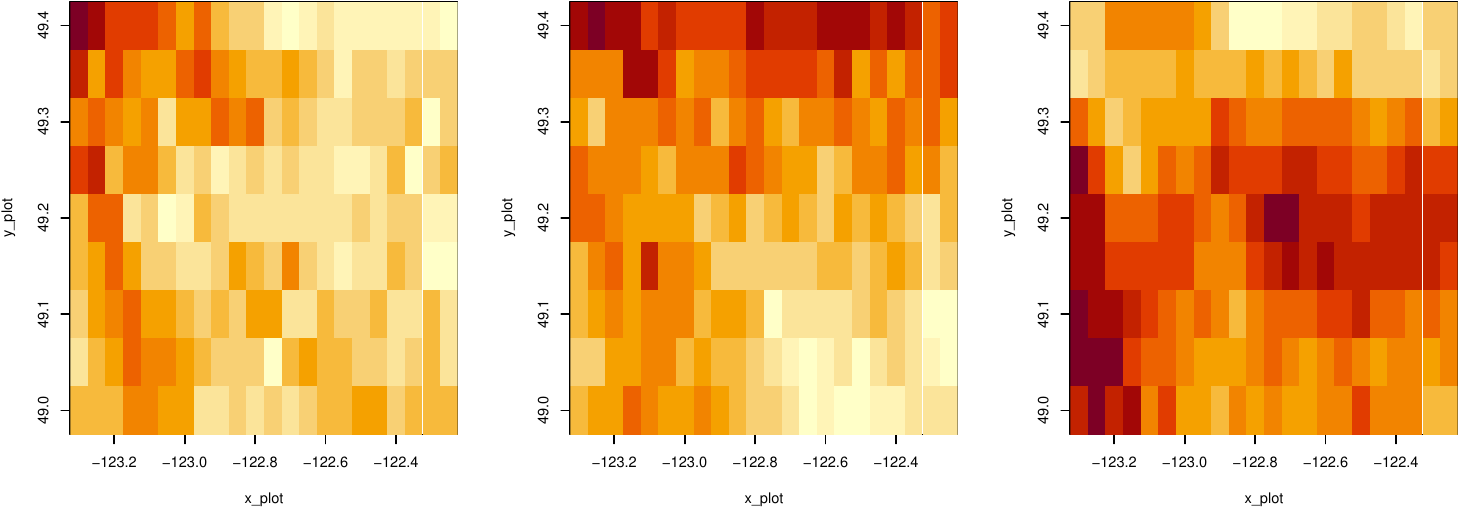} 

}

\caption[Three simulated realizations of GPs based on the bijective deformation model]{Three simulated realizations of GPs based on the bijective deformation model.}\label{fig:simulate2}
\end{figure}
\end{CodeChunk}

Note that simulations are based on a Cholesky decomposition with
pivoting of the model-based variance-covariance matrix from the fitted
model. The pivoting allows for rank-deficient GPs to be simulated.
However, this is likely to be very slow and memory-hungry for
high-dimensional simulations.

\section{Censored data}

\subsection{Correlation and covariance matrix estimation}

Sometimes we might want to estimate a GP from censored data. When only
interested in the tail of some data, instead of modeling all data to
estimate a distribution, only the values of exceedances of threshold can
be considered, and the values of non-exceedances treated as censored.
\cite{bor-ct} introduced the multivariate Gaussian tail model in which a
generalized Pareto model represented exceedances of a threshold and then
a probability integral transformation gave Gaussian data. These tails
were then modeled as multivariate Gaussian, which was estimated using a
pairwise censored likelihood approach; see \cite{ledford1997}, for
example.

The \pkg{deform} package includes three functions related to censored
data.

\begin{Code}
  fitcenmvn(x, left)
  cencor(x, left, scale)
  cencov(x, left, scale)
\end{Code}

Function \code{fitcenmvn()} takes \(n \times m\) matrices \code{x} and
\code{left} and estimates a \(m \times m\) variance-covariance matrix by
fitting bivariate Gaussian distributions to each pairwise combination of
columns of \code{x} by assuming that the \((t, i)\)th entry in \code{x}
is left-censored at the \((t, i)\)th entry in \code{left}, for
\(t = 1, \ldots, n\) and \(i = 1, \ldots, m\). Functions \code{cencor()}
and \code{cencov()} return correlation and covariance matrices,
respectively, which are scaled by the mean estimate if
\code{scale = TRUE}; otherwise \code{cencor()} and \code{cencov()}
assume a zero mean.

The functions above operate by first fitting marginal censored Gaussian
distributions, i.e.~assuming for data
\(y_1(\boldsymbol{x}), \ldots, y_n(\boldsymbol{x})\) a likelihood of the
form \[
L(\mu(\boldsymbol{x}), \tau(\boldsymbol{x})) = \prod_{t = 1}^n h(y_t(\boldsymbol{x}); u_t(\boldsymbol{x}), \mu(\boldsymbol{x}), \tau(\boldsymbol{x})),
\] where \[
h(y; u, \mu, \tau) = \left\{\begin{array}{ll} \Phi_1(u; \mu, \tau) & \text{if } y \leq u,\\ \phi_1(y; \mu, \tau) & \text{if } y > u,\end{array}\right.
\] where \(\Phi_1(\, ; \mu, \tau^2)\) and \(\phi_1(\, ; \mu, \tau^2)\)
denote the mean \(\mu\) variance \(\tau^2\) univariate Gaussian
cumulative distribution (cdf) and probability density (pdf) functions,
respectively. The fitted marginal Gaussian distributions are then used
to give
\(w_t(\boldsymbol{x}) = [y_t(\boldsymbol{x}) - \hat \mu(\boldsymbol{x})] / \hat \tau(\boldsymbol{x})\).
Pairwise combinations are modeled as unit bivariate Gaussian, so that
thecorrelation between \(W_t(\boldsymbol{x})\) and
\(W_t(\boldsymbol{x}')\), \(\rho(\boldsymbol{x}, \boldsymbol{x}')\), is
estimated by maximizing the likelihood
\[L\big(\rho(\boldsymbol{x}, \boldsymbol{x}')\big) = \prod_{t=1}^n f\big(w_t(\boldsymbol{x}), w_t(\boldsymbol{x}'); u_t(\boldsymbol{x}), u_t(\boldsymbol{x}'), \rho(\boldsymbol{x}, \boldsymbol{x}')\big)\]
where \(f\big(w, w'; u, u', \rho\big)\) is given by
\begin{equation*}\left\{\begin{array}{lll} \Phi_2\big(u, u'; \rho) & \text{if} & w \leq u, w' \leq u',\\ \phi_1(w'; 0, 1) \Phi_1(u; \rho w', 1 - \rho^2)  & \text{if} & w \leq u, w' > u',\\ \phi_1(w; 0, 1) \Phi_1(u'; \rho w, 1 - \rho^2)  & \text{if} & w > u, w' \leq u',\\ \phi_2\big(w, w'; \rho) & \text{if} & w > u, w' > u',\end{array}\right. \end{equation*}
and \(\phi_2(\, , \,; \rho)\) and \(\Phi_2(\, , \,; \rho)\) denote the
bivariate unit Gaussian cdf and pdf, respectively. The bivariate
Gaussian cdf is approximated using an \pkg{Rcpp} implementation of Alan
Gentz's MVTDST set of Fortran subroutines
(\url{https://www.math.wsu.edu/faculty/genz/software/software.html}).

\subsection{A simple stochastic rainfall model}

Here we demonstrate a simple use of the left-censored multivariate
Gaussian distribution by producing a basic stochastic weather generator
of daily rainfall. This uses the rainfall data of \cite{cooley2007} by
modeling the tail of rainfall measurements from 1st April to 31st
October in the years 1990 to 2019 over 10mm as multivariate Gaussian.
This model is for demonstration purposes only: a better approach might
be to model all rainfall amounts using a zero-inflated gamma or Weibull
distribution and then use a probability integral transformation so that
the multivariate Gaussian tail model can be used.

The following loads the data (which are given as datasets \code{COprcp}
and \code{COprcp_meta} in package \pkg{evgam}), combines the rainfall
data with its meta data, and then forms a matrix of the data in which
each column corresponds to station and each row to a daily rainfall
total.

\begin{CodeChunk}
\begin{CodeInput}
R> library(evgam)
R> data(COprcp)
R> COprcp <- cbind(COprcp, COprcp_meta[COprcp$meta_row,])
R> COprcp <- subset(COprcp, as.integer(substr(date, 6, 7)) %in% 4:10)
R> tms <- as.factor(COprcp$date)
R> locs <- as.factor(COprcp$meta)
R> COprcp_mat <- matrix(NA, length(levels(tms)), length(levels(locs)))
R> COprcp_mat[cbind(as.integer(tms), as.integer(locs))] <- COprcp$prcp
\end{CodeInput}
\end{CodeChunk}

We then fit a bivariate Gaussian distribution to rainfall measurements
for each pair of stations, treating the measurements as left-censored at
10mm.

\begin{CodeChunk}
\begin{CodeInput}
R> thresh <- 10
R> thresh_mat <- matrix(thresh, nrow(COprcp_mat), ncol(COprcp_mat))
R> mS <- fitcenmvn(COprcp_mat, thresh_mat)
\end{CodeInput}
\end{CodeChunk}

The pairwise covariance matrix estimate can then act as
\(\boldsymbol{V}\) in equation \eqref{gplik}. This is supplied below to
\code{expand()} and a rank-12 dimension expansion model is then fitted.

\begin{CodeChunk}
\begin{CodeInput}
R> n0 <- round(ncol(COprcp_mat) * mean(!is.na(COprcp_mat)))
R> m0 <- expand(as.matrix(COprcp_meta[, c('lon', 'lat')]), mS[[2]], n0, k = 12)
\end{CodeInput}
\end{CodeChunk}

The following visualises the dimension expansion

\begin{CodeChunk}
\begin{CodeInput}
R> par(mfrow = c(1, 3))
R> plot(m0)
\end{CodeInput}
\begin{figure}

{\centering \includegraphics{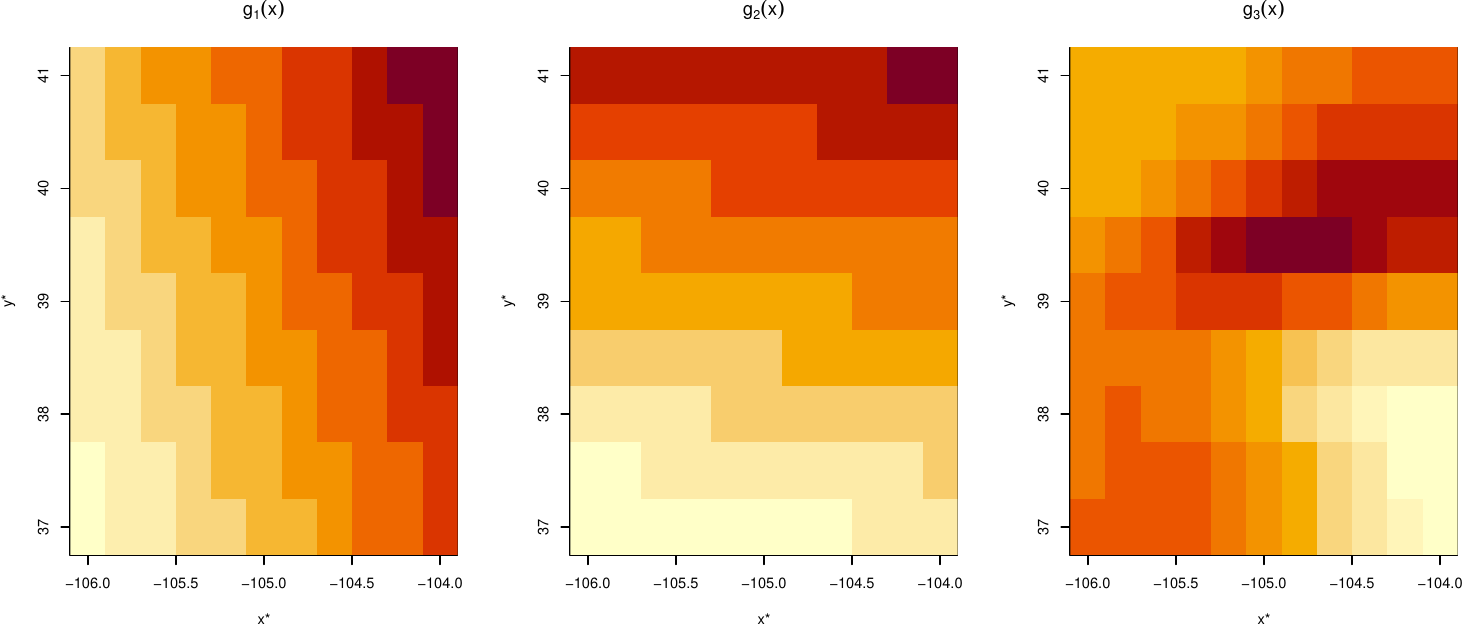} 

}

\caption[Dimension expansion model representation for censored data]{Dimension expansion model representation for censored data.}\label{fig:prcp4}
\end{figure}
\end{CodeChunk}

and then we can also view its variogram, which appears to show an
adequate fit.

\begin{CodeChunk}
\begin{CodeInput}
R> variogram(m0)
\end{CodeInput}
\begin{figure}

{\centering \includegraphics[width=0.65\linewidth]{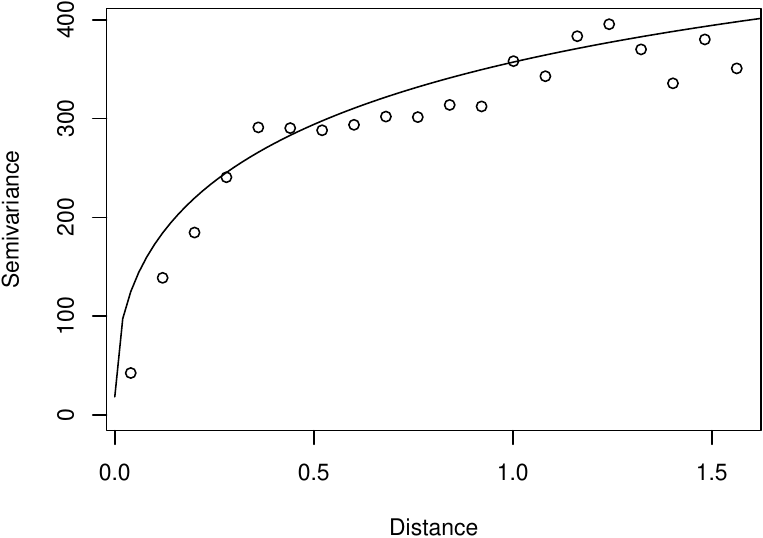} 

}

\caption[Variogram for censored data]{Variogram for censored data.}\label{fig:prcp5}
\end{figure}
\end{CodeChunk}

Finally, we can use the fitted model to simulate rainfall amounts over
10mm. The following simulates ten realizations of daily rainfall at each
station and then shows the amounts for the first six stations, with any
simulated amounts below 10mm shown as \code{NA}.

\begin{CodeChunk}
\begin{CodeInput}
R> set.seed(5)
R> w_sim <- mS[[1]] + simulate(m0, nsim = 12)
R> y_sim <- replace(w_sim, w_sim <= thresh, NA)
R> head(y_sim)
\end{CodeInput}
\begin{CodeOutput}
     [,1] [,2] [,3] [,4] [,5] [,6]     [,7]     [,8]     [,9] [,10] [,11] [,12]
[1,]   NA   NA   NA   NA   NA   NA 11.39131 16.85773       NA    NA    NA    NA
[2,]   NA   NA   NA   NA   NA   NA       NA       NA 46.18505    NA    NA    NA
[3,]   NA   NA   NA   NA   NA   NA       NA       NA       NA    NA    NA    NA
[4,]   NA   NA   NA   NA   NA   NA       NA       NA       NA    NA    NA    NA
[5,]   NA   NA   NA   NA   NA   NA       NA       NA       NA    NA    NA    NA
[6,]   NA   NA   NA   NA   NA   NA 11.15420       NA       NA    NA    NA    NA
\end{CodeOutput}
\end{CodeChunk}

The above example is a simple demonstration based on a fixed set of
stations. More generally, the multivariate Gaussian tail model of
\cite{bor-ct} can be coupled with continuous models for threshold
exceedances, such as in \cite{y2019}, to give a continuous probability
integral transformation to unit Gaussian, and then a nonstationary GP
fitted with \pkg{deform}. This would allow simulations for an entire
domain, as opposed to a fixed set of sites.

\section{Summary}

The \proglang{R} package \pkg{deform} is designed to give functions for
nonstationary GP modeling and subsequent visualization of spatial data.
It allows users to fit GPs with a conventional anisotropic stationary
dependence structure, and also with a nonstationary structure, based on
the methods of spatial deformation and dimension expansion. Deformations
and latent dimensions are achieved using thin plate regression splines
to take coordinates in \(G\)-space and give coordinates in \(D\)-space
such that covariance becomes closer to isotropic. Having fitted a model,
users can view spatial deformations and dimension expansions. For
example, points brought closer together in \(D\)-space have stronger
dependence, given their positions in \(G\)-space. Users can also use
variograms to assess the fit a \pkg{deform} model, which allows the
model-based representation of covariance to be compared against its
empirical counterpart. Predictions of points in \(D\)-space for
arbitrary points in \(G\)-space are possible, and simulations of
realizations of nonstationary GPs are also possible, given arbitrary
points in \(G\)-space.

The models in \pkg{deform} are likely to be very time-consuming to fit
for high dimensional data due to various \(O(m^3)\) calculations, such
as Cholesky decompositions. It is hoped that updates can be brought to
\pkg{deform} that facilitate the modeling of such data, in particular
gridded climate model output. Facilitating high-dimensional simulations
is likely to be implemented prior to high-dimensional model fitting. The
\pkg{deform} package has been introduced by considering
\(\boldsymbol{x}\) defined in terms of longitude and latitude. Although
\(G\)-space must be two-dimensional, it does not need to be defined
geographically: see \cite{cooley2007} for the notion of `climate space'.
Future developments may allow the possibility for latent dimensions in
the dimension expansion approach to be defined in terms of covariates.
For example, adding a latent dimension that is a function of elevation
may help bring isotropy for some environmental phenomena.

Uncertainty estimation in \(\boldsymbol{g}\) currently does not allow
for uncertainty in \(\boldsymbol{\lambda}\). This could be achieved
using the method of \citet[\S 6.11.1]{wood-book}, or the more general
method of \cite{rue2009}. Finite-differencing can be used for some of
the more complex derivatives.

\section*{Computational details}

The results in this paper were obtained using \proglang{R} 4.3.1 with
the \pkg{deform} 1.0.1 package. \proglang{R} itself and \pkg{deform} are
available from the Comprehensive \proglang{R} Archive Network (CRAN) at
\url{https://CRAN.R-project.org/}.

\bibliography{deform.bib}

\begin{appendix}

\section{Sensitivity to parameters that control bijectivity} \label{pensens}

The following shows the effects of changing $\delta$ and $\epsilon$, which control the folding penalty, and are defined in equation \eqref{foldpen}. Changes by one order of magnitude in either direction are considered. 

\begin{CodeChunk}
\begin{CodeInput}
R> bij_pars <- cbind(mult = c(1e2, 1e4, 1e3, 1e3), scl = c(1, 1, .1, 10))
R> par(mfrow = c(2, 2), mar = c(2, 2, 3, 1))
R> for (i in 1:4) {
+   args_i <- list(mult = bij_pars[i, 1], scl = bij_pars[i, 2])
+   mod_i <- deform(solar, bijective = TRUE, bijective.args = args_i)
+   plot(mod_i, xp = x_plot, yp = y_plot, asp = 1)
+   title(paste('mult =', args_i$mult, ', ', 'scl =', args_i$scl))
+ }
\end{CodeInput}
\begin{figure}

{\centering \includegraphics[width=0.8\linewidth]{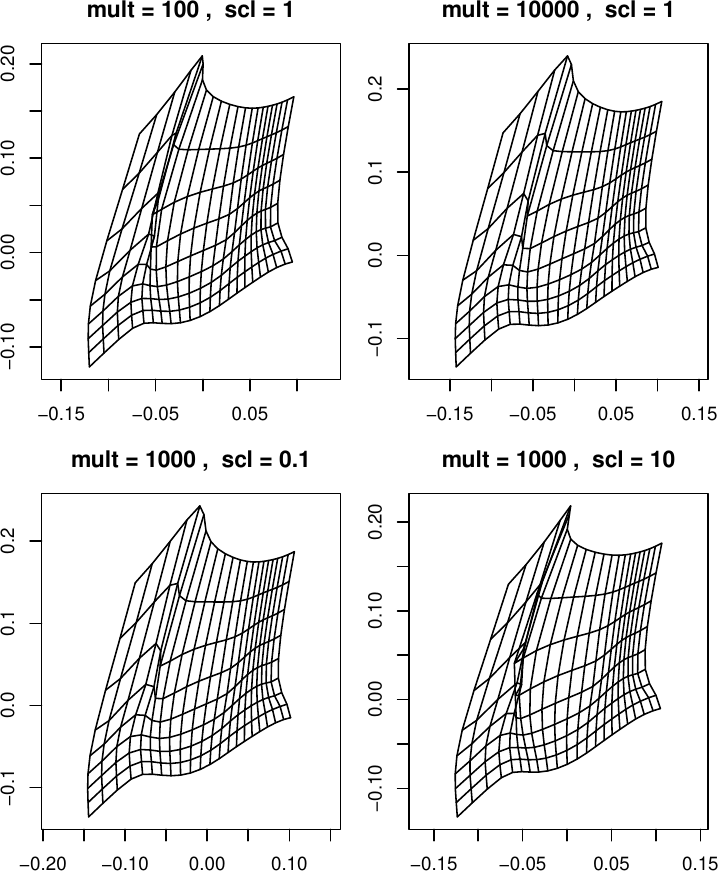} 

}

\caption{Sensitivity to changes in bijectivity arguments \code{mult} and \code{scl}, wihch are related to $\delta$ and $\epsilon$, as detailed in \S\ref{S:solar:biject}.}\label{fig:bijective2}
\end{figure}
\end{CodeChunk}

In summary, too small $\delta$ or too high $\epsilon$ (which shrinks areas) fails to impose enough of a penalty to ensure bijectivity. Therefore if \code{deform()}'s default values fail to bring bijectivity, $\delta$ should be increased and/or $\epsilon$ should be decreased. Users are advised to try increasing $\delta$ first. 

\end{appendix}

\end{document}